\RequirePackage{lineno}

\documentclass[floatfix,superscriptaddress,a4paper,
               showpacs,showkeys,nofootinbib,preprint]{revtex4}

\usepackage{amsfonts,amsmath,soul}
\usepackage{amsmath}

\usepackage{epsfig}
\usepackage{latexsym}
\usepackage{xspace}
\usepackage{relsize}

\usepackage{inputenc}
\usepackage{indentfirst}
\usepackage{enumerate}
\usepackage{color}
\usepackage{epstopdf}
\usepackage{colortbl}
\usepackage{blindtext}

\usepackage{amsmath}
\usepackage{amssymb}
\usepackage[english]{babel}
\usepackage{hyperref}

\usepackage{soul}

\graphicspath{{plots/}}

\newcommand{\eV}{\ensuremath{\mbox{e\kern-0.1em V}}\xspace}
\newcommand{\GeV}{\ensuremath{\mbox{Ge\kern-0.1em V}}\xspace}
\newcommand{\MeV}{\ensuremath{\mbox{Me\kern-0.1em V}}\xspace}
\newcommand{\GeVc}{\ensuremath{\mbox{Ge\kern-0.1em V}\!/\!c}\xspace}
\newcommand{\GeVcc}{\ensuremath{\mbox{Ge\kern-0.1em V}\!/\!c^2}\xspace}
\newcommand{\AGeV}{\ensuremath{A\,\mbox{Ge\kern-0.1em V}}\xspace}
\newcommand{\AGeVc}{\ensuremath{A\,\mbox{Ge\kern-0.1em V}\!/\!c}\xspace}
\newcommand{\MeVc}{\ensuremath{\mbox{Me\kern-0.1em V}/c}\xspace}

\newcommand{\X}{{\mathcal{X}}\xspace}
\newcommand{\Y}{{\mathcal{Y}}\xspace}
\newcommand{\B}{{\mathcal{B}}\xspace}

\newcommand{\eq}[1]{\begin{align} #1 \end{align}}

\begin{document}

\title{
Equilibration and locality
}

\author{M. Gazdzicki}
\affiliation{Goethe-University Frankfurt am Main, Germany}
\affiliation{Jan Kochanowski University, Kielce, Poland}
\author{M. I. Gorenstein}
\affiliation{Bogolyubov Institute for Theoretical Physics, Kyiv, Ukraine}
\author{I. Pidhurskyi}
\affiliation{Goethe-University Frankfurt am Main, Germany}
\author{O. Savchuk}
\affiliation{Bogolyubov Institute for Theoretical Physics, Kyiv, Ukraine}
\affiliation{Frankfurt Institute for Advanced Studies Frankfurt am Main, Germany}
\author{L. Tinti}
\affiliation{Goethe-University Frankfurt am Main, Germany}
\affiliation{Jan Kochanowski University, Kielce, Poland}

\begin{abstract}


Experiments
motivated by predictions of quantum mechanics
indicate non-trivial correlations between spacelike-separated measurements.
The phenomenon is referred to as
a violation of strong-locality and, after Einstein, called \emph{ghostly action at a distance}.
An intriguing and previously unasked question is how    
the evolution of an assembly of particles to equilibrium-state  
relates to strong-locality.
More specifically, whether, with this respect, indistinguishable particles differ from distinguishable ones.

To address the question, we introduce a Markov-chain based framework 
over a finite set of microstates.
For the first time, we formulate conditions needed to
obey the particle transport- and strong-locality 
for indistinguishable particles. 

Models which obey transport-locality and
lead to equilibrium-state are considered.
We show that it is possible to construct models obeying and violating
strong-locality both for indistinguishable particles and for distinguishable ones.
However, we find that 
only for distinguishable particles strongly-local evolution to equilibrium
is possible without breaking the microstate-symmetry.
This is the strongest symmetry one can impose and leads to the shortest equilibration time.

We hope that the results presented here may provide a new perspective on a violation of strong-locality, and the developed framework will help in future studies. 
Specifically they may help to interpret results on high-energy nuclear collisions indicating a fast equilibration of indistinguishable particles.
\end{abstract}


\maketitle


\newpage

\section{Introduction}
\label{sec:introduction}

It is commonly accepted that experimental results
indicate non-trivial correlations between spacelike-separated measurements.
The effect is deeply rooted in quantum mechanics~\cite{Einstein:1935rr,Bell:1964kc, Horodecki:2009zz,Alford:2015xpa} and
is referred to as  a violation of strong-locality or local-causality. 
On the other hand, quantum statistics, observed in many experiments, can be derived using statistical methods for indistinguishable particles
- the particles of quantum mechanics~\cite{Fierz:1939zz,Pauli:1940zz,reif1965}.
We address here a previously unasked question -  how is strong-locality related to equilibration?
More specifically, whether, concerning this question, indistinguishable particles differ from distinguishable ones.
The question is discussed within a Markov chain framework
over a finite set of microstates.

\begin{figure}[ht]

\includegraphics[width=0.90\textwidth]{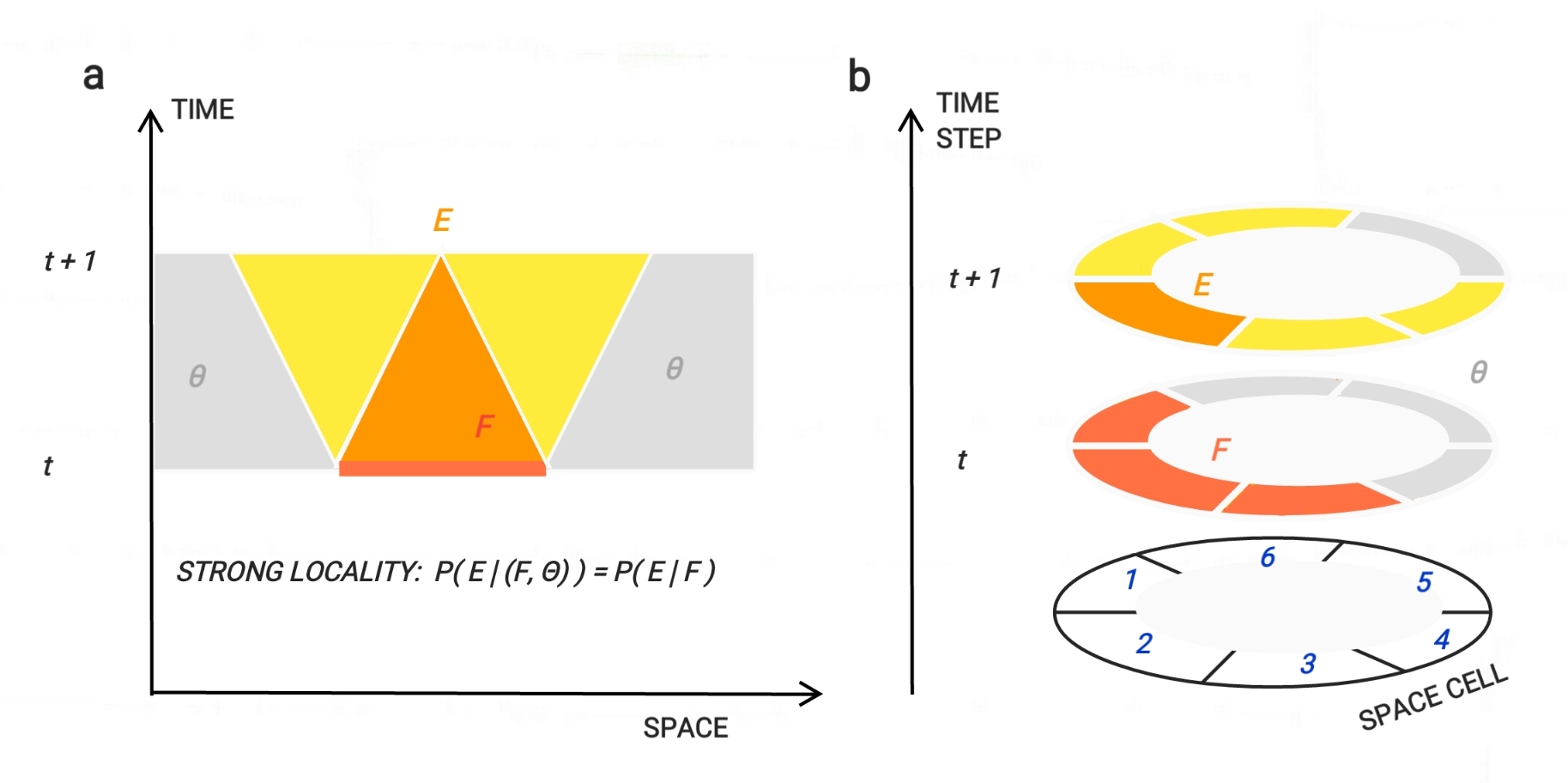} 
    \caption{
    Sketches illustrating the principle of strong-locality:
    (a) the illustration in continuous space-time and
    (b) the illustration within the Cell Model with six cells, $\Delta=1$ and periodic boundary conditions, see text for details.
}
\label{fig:strong_locality}

\end{figure}

The principle of 
strong-locality states that an event $E$ depends only on events $F$ in the event's past light cone~\cite{Alford:2015xpa}; for illustration, see a sketch in Fig.~\ref{fig:strong_locality}~(a).
Thus,
whatever the initial (at time $t$) microstates having $F$, 
the probability of observing the event $E$ (at time $t+1$) is the same. 
Referring to Fig.~\ref{fig:strong_locality}, it is independent of
events $\theta$ located in the grey regions.
However, the probability may depend on events 
in the yellow and orange regions of the plot. It is because their 
light cones between $t$ and
$t+1$ overlap with the light cone of $E$.
The strong-locality principle embeds a weaker but more intuitive property. The conserved quantities cannot be transported faster than the speed of light in the vacuum. 
Experimental measurements~\cite{Aspect:1982fx,Aspect:1981nv,Aspect:1981zz,Hensen:2015ccp} give evidence of a violation of strong-locality in nature.

Numerous experiments show
that isolated systems of interacting particles evolve from an arbitrary initial microstate to a steady-state - the state whose macroscopic properties are independent
of time. Frequently steady-state properties agree
with predictions of quantum statistics for the maximum entropy state - 
the equilibrium-state.

Many physics models are based on the mathematical framework introduced by Markov to
describe stochastic processes - Markov chains.
Among the best-known ones is the Einstein-Smoluchowski description~\cite{Einstein:1926,Smoluchowski:1923} of the Brownian
motion. It motivated Nelson~\cite{Nelson:1966sp,Nelson:2012} to 
interpret quantum-mechanical phenomena
within stochastic mechanics~\cite{Nelson:2012} based on Markov chains.
It is disputed~\cite{Gillespie:1994,PhysRevA513445,Garbaczewski:1996dv} whether transitions of Markov chains may resemble measurements of
quantum physics.

In this paper
a Markov chain framework for studying the time evolution of the assembly of particles is introduced. It allows addressing questions on the relation between 
the equilibrium-state and strong-locality in a simple indeterministic approach.
Its basic parameters 
are transition probabilities between microstates determining a configuration of all particles
at a given time. 
Isolated systems with the conserved number of
distinguishable and indistinguishable particles are considered.
We summarise the well-known constraints for the transition matrix regarding space-time reflection and translation invariance and conditions needed to reach  the equilibrium-state. 
For the first time, we introduce constraints granting the particle transport and 
strong-locality for indistinguishable particles.

Models which obey particle transport-locality 
and lead to the equilibrium-state are considered.
We show that it is possible to construct models obeying and violating
strong-locality both for indistinguishable particles and for distinguishable ones.
To uncover differences between the two types of particles, 
we introduce the microstate-symmetry - the strongest symmetry that implies
the equiprobability of allowed transitions.
This leads us to an answer for the addressed question - 
without breaking the microstate-symmetry,
the strongly-local evolution to equilibrium
is possible only for distinguishable particles.

\section{Model}
\label{sec:model}

This section firstly introduces the Symmetric Cell Models  - the Markov chains,  
which obey space-time symmetries~(\ref{sec:scm}).
Within the models,  indistinguishable and distinguishable particles are considered.
Then requirements on the transition matrix needed to reach 
equilibrium-state~(\ref{sec:equilibrium}) are summarised.
The presentation is closed by giving the conditions for
transport-locality~(\ref{sec:tl}) and strong-locality~(\ref{sec:sl}) of the system evolution.

\subsection{Symmetric Cell Models}
\label{sec:scm}

The simplest way to address the locality of evolution and approach to the equilibrium-state in a dynamic model is to consider a  1+1 dimensional discrete-time Markov chain with a conserved number of particles. 

The following is postulated:
\begin{enumerate}[(a)]
\item
\emph{Space}: Space is assumed to be a vector of $V$ discrete cells $(v_1, v_2, \ldots, v_V)$ arranged in 
a one-dimensional ring. The periodic boundary conditions are needed for the space translation invariance (see below).
For illustration a sketch for $V = 6$ is presented in Fig.~\ref{fig:strong_locality}~(b).

\item
\emph{Particles}: 
The total number of particles $N$ is constant.
For simplicity, only spin-zero particles are considered.
The number of particles in a cell can vary between 0 and $N$.
Indistinguishable and distinguishable particles are considered.

\item
\emph{Microstates}: The microstates of the system are characterised either by numbers of particles in cells for indistinguishable particles or by an arrangement of distinguishable particles. 
A microstate $X$ of indistinguishable particles is given by the sequence of  particle multiplicities in cells,

\eq{
\label{eq:X_ind} 
X = (n_1^X, n_2^X, \ldots  ,n_V^X)~,
}
where $\sum\limits_{i=1}^{V}n_i^X =  N$.
The total number of microstates (see, e.g. Ref.~\cite{Gazdzicki:2020cyp}) is
\eq{
\label{eq:W_ind}
W_{\rm ind}(N,V) = \frac{(N+V-1)!} {N!(V-1)!}~.
}

A microstate of distinguishable particles is denoted by $\X$, and it is given by the
the sequence of cell numbers of particles for all $N$ particles

\eq{
\label{eq:X_dist} 
\X = (v_1^X, v_2^X, \ldots  ,v_N^X)~.
}
The total number of  microstates for distinguishable particles reads~~\cite{Gazdzicki:2020cyp}

\eq{
\label{eq:W_dist}
W_{\rm dist}(N,V) = V^N~.
}
The above expression corresponds to the 
well-known result~\cite{bookFeller} for the number of different arrangements of $N$ labelled balls (distinguishable particles) among $V$ labelled boxes (cells).

\item
\emph{Time, time steps and transitions}: 
The system's evolution in \textit{time} $\tau$ is assumed to be discrete.
During the evolution, \emph{transitions} between its microstates occur.
The \emph{time steps} $t$ at which microstates of the system 
appear are equally distant in time, $\tau = \delta \tau \cdot t$,
where $\delta \tau$ is a time interval between to consecutive time steps.
The transition probability from an initial microstate at $t$ to a final
microstate at $t+1$ is assumed to depend only on the initial microstate.
It is independent of microstates preceding the initial microstate  
and the time step $t$.
The transition probability for indistinguishable particles is denoted by 

\eq{\label{eq:TPM_ind}
B(X \to Y|X)~,
}
whereas for distinguishable ones, it is

\eq{\label{eq:TPM_dist}
\B(\X \to \Y|\X)~.
}
Unlike otherwise stated, explicit expressions are given only for
indistinguishable particles. The corresponding expressions for distinguishable particles
can be obtained by  substituting $X$ by $\X$,  $Y$ by $\Y$ and  $B$ by  $\B$.

Since the matrix represents probabilities, each of its elements must be non-negative and no larger than one, and the sum of the probabilities of all transitions from a given initial microstate must be equal to one
\label{unit}
\eq{
B(X\to Y|X)\in[0,1], \quad \forall X,Y;  \qquad \sum_Y B(X \to Y|X)=1, \quad \forall X~.
}

Let $\tilde{P}(X, t)$ denotes a probability to find a microstate $X$ in the ensemble of systems at $t$. By definition $\sum_X \tilde{P}(X,t) = 1$ for all $t$.
Then the probability of finding a microstate $Y$ at $t+1$ is

\eq{
\label{eq:PY}
\tilde{P}(Y, t+1) = \sum_X B(X \to Y|X)~\tilde{P}(X, t) ~.
}
\end{enumerate}

\vspace{0.2cm}
The above assumptions define a class of Cell Models~\mbox{\cite{Gazdzicki:2017rfe,Gazdzicki:2020cyp}} -  discrete-time Markov chains on a finite set of microstates~\cite{ergodic}.
The definition of a Cell Model is completed by specifying the model transition matrix.
The matrix encapsulates introduced constraints - space-time symmetries, particle number conservation law, transport-locality - and dynamics.

Symmetric Cell Models are the Cell Models which obey the space-time symmetries: 

\begin{enumerate}[(i)]
 \item
    \emph{Space-translation symmetry}: Given a translation $T$ of particles in the cell space, the transition probability from a microstate $X$ to a microstate $Y$ is the same as the transition probability between translated microstates $T(X) \to T(Y)$:
    
    \eq{
    B(X \to Y | X) = B(~T(X) \to T(Y) | T(X)~)~.
    }

    \item
    \emph{Space-reversal symmetry}: Given a reflection $R$ of particles with respect to an axis in the cell space, the probability of any transition is equal to the probability of the transition between the reflected microstates:
    
    \eq{
    B(X \to Y | X) = B(~R(X) \to R(Y) | R(X)~)~.
    }

    \item
    \emph{Time-reversal symmetry}:
    The time-reversal symmetry for ergodic systems (see below) is obeyed if and only if,
    for every recurring time sequence of microstates, $(X_1, X_2, \ldots, X_m, X_1)$, the probability of visiting the microstates in the original order or the reversed one is the same.
    Expressed in the transition matrix elements, the  condition reads:
    
    \begin{equation}
    \label{eq:Kolmogorov}
    \begin{split}
    & B(X_1 \to X_2 | X_1) \cdot B(X_2 \to X_3 | X_2) \cdots 
    B(X_{m-1} \to X_m | X_{m-1}) \cdot B(X_m \to X_1 | X_m) = \\
    & B(X_1 \to X_m | X_1) \cdot B(X_m \to X_{m-1} | X_m) \cdots 
    B(X_3 \to X_2 | X_3) \cdot B(X_2 \to X_1 | X_2)~,
   \end{split}
   \end{equation}
and it is known as the Kolmogorov cycling condition~\cite{Kolmogorov_1,Kolmogorov_2}.
    
\end{enumerate}

\subsection{Steady- and equilibrium-states}
\label{sec:equilibrium}

\emph{Steady-state} of a Markov chain corresponds to the probability distribution
$\tilde{P}(X, t)$ of finding a microstate $X$ at $t$ that is independent of $t$.
This distribution is denoted by
$\pi_S(X)$ and it is the eigenstate of the transition matrix corresponding to the eigenvalue $1$. 

\emph{Equilibrium state} is a steady-state, which corresponds to  the probability distribution $\pi_{\rm eq}$ maximising the entropy. In the absence of any further  constraints than the accessible microstates, the maximum entropy of a system  
corresponds to the equiprobable distribution. In other words, $\pi_{\rm eq}=1/W$~, where $W = W_{\rm ind}$ and $W=W_{\rm dist}$ for indistinguishable and distinguishable particles~\cite{Gazdzicki:2017rfe}, respectively. 

Properties of the equilibrium-state were studied within the Cell Model in
Ref.~\cite{Gazdzicki:2020cyp}, where it was shown that a distribution 
of particle multiplicity in a cell resembles the Bose-Einstein distribution 
for indistinguishable particles
and the Poisson distribution for distinguishable ones.

The two  sufficient conditions to reach an equilibrium-state, regardless of the initial distribution, are ergodicity and transition-matrix symmetry~\cite{ergodic}.
Ergodicity is necessary and sufficient to ensure that the transition matrix has a unique steady-state~\cite{ergodicity_1,ergodic}, and that the probability distribution asymptotically approaches it, regardless of the initial microstate.
Time reversibility~\eqref{eq:Kolmogorov}, which is assumed in the Symmetric Cell Models, is equivalent to the detailed balance condition~\cite{Kolmogorov_1,Kolmogorov_2}:

\eq{
 \pi_S(X) \cdot B(X\to Y|X)=\pi_S(Y) \cdot B(Y\to X|Y)~.
}
Therefore a symmetric transition matrix (the matrix equal to its transpose) leads to the equilibrium steady-state, $\pi_S=\pi_{\rm eq}$.

We remind here the two conditions to fulfil for a finite discrete Markov chain to be ergodic~\cite{ergodic}. These are:

\begin{enumerate}[(a)]
 \item
 \emph{Irreducibility}: Each microstate is reachable from any other microstate by a sequence of transitions with non-vanishing probability.
 \item
 \emph{Aperiodicity}: The maximum common divisor of the number of transitions of each sequence linking one microstate to itself is one.
\end{enumerate}
Ergodicity of discussed here matrices is tested in Appendix~\ref{app:a}.

We call a Symmetric Cell Model \emph{equilibrating} if it is ergodic and it reaches the equilibrium-state. This is  equivalent to the model having an ergodic and symmetric transition matrix.
Models leading to the equilibrium-state are discussed in the paper and denoted EQ$+$. 
An interesting example of a non-equilibrating model (EQ$-$) for indistinguishable particles is given in Appendix~\ref{app:b} and discussed in Appendix~\ref{app:c}.
The assumption of statistical and local (cell-by-cell) redistribution of cell multiplicities, unlike naively expected,
leads to a non-equilibrium (non-statistical) steady-state.

\subsection{Transport-locality}
\label{sec:tl}

Let us postulate that in a single time step, particles (in general, conserved quantities) can be displaced by no more than $\Delta$ cells.
We call this requirement \emph{transport-locality}.
We remind  that transport-locality is necessary but not sufficient to obey strong-locality~\cite{Alford:2015xpa}.

For distinguishable particles, the transport-locality requirement reduces to a requirement of each  particle moving by no more than $\Delta$ cells during a single time step. In physics, this corresponds to
particle velocity being limited by the speed of light in the vacuum. 

For indistinguishable particles, particle's trajectories and 
velocities are undefined.
Then the transport-locality implies the following.
During a single time step, the particle number in any interval of cells cannot be
transported beyond an interval by $\Delta$ cells longer on the left and right.
And it cannot be squeezed to an interval by $\Delta$ cells shorter on the left and right.
This provides two transport-locality inequalities:

\eq{
\label{eq:local_leq}
\sum_{j=i}^{i+k}n_j^X \leq \sum_{l=i-\Delta}^{i+k+\Delta} n_l^Y~, \\ 
\nonumber
\sum_{l=i}^{i+k}n_l^Y \leq \sum_{j=i-\Delta}^{i+k+\Delta} n_j^X~,
}
where $k = 0, 1, \ldots $ and $n_j^X$, $n_l^Y$ are particle numbers in cells $j$, $l$ of
initial ($X$) and final ($Y$) microstates, respectively.

We note that if $\delta \tau$ and $\Delta$ depended on $t$,
different transitions would be allowed or excluded by transport-locality at different time steps. Thus the transition matrix would depend on the time step. The resulting model would not be a Markov chain any longer.
Consequently we assume that the time interval between
two consecutive time steps, $\delta \tau$, is independent of the time step and thus
$\Delta = const$. 
To simplify the presentation, we discuss models with $\Delta =1$.

In the following transitions obeying transport-locality are referred to as transport-local
transitions and the models allowing only the transport-local transitions will be denoted TL$+$.
Models that violate transport-locality  are denoted TL$-$.

\subsection{Strong-locality}
\label{sec:sl}
According to the strong-locality principle, the probability of an event $E$ at $t+1$ can be influenced only by events within its past light cone.  
Thus, $E$ must be independent of events outside its light cone when possible common-past correlations  are removed.
Referring to Fig.~\ref{fig:strong_locality}, given an initial configuration at $t$, any event $E$ at $t+1$ can depend only on the configurations $F$ of the system at time $t$, in the dark orange line in the plot~(a). 
The event $E$  is independent of all events $\theta$ in grey areas
of Fig.~\ref{fig:strong_locality}:

\eq{\label{eq:strong_locality}
P( E~|~(F, \theta)~) = P( E~|~F )
}
for all allowed events $F$ and $\theta$.
Note that events in the yellow and orange regions  can be correlated with $E$ because possible 
common-past correlations are not removed.

Violation of strong-locality can be quantified by introducing the strong-locality parameter:

\eq{\label{eq:Rstrong}
\nu_{SL} \equiv \mathlarger{\sum}_{E}~ \mathlarger{\sum}_{F}~ \mathlarger{\sum}_{\theta} \left( P( E~|~(F, \theta)~) - \overline{P(E~|~(F,\theta)~)} \right)^2~,
}
where $\overline{P( E~|~(F,\theta)~)}$ is obtained by averaging $P(E~|~(F,\theta)~)$ over all possible $\theta$ for given $E$ and $F$. Obviously, for an evolution obeying 
strong-locality $\overline{P( E~|~(F,\theta)~)}= P( E~|~(F,\theta)~) = P( E~|~F )$, and thus $\nu_{SL} = 0$. 
The matrices obeying strong-locality will be denoted SL$+$ and those violating it
SL$-$.

Figure~\ref{fig:strong_locality}~(b) shows an example sketch for the  Cell Model with $V=6$ cells and $\Delta = 1$.
In this example the test of strong-locality can be done by setting
$E = n_2^{(t+1)}$, $F = (n_1^{(t)}, n_2^{(t)}, n_3^{(t)} )$ and
$\theta = ( n_5^{(t+1)},~~ n_4^{(t)}, n_5^{(t)}, n_6^{(t)} )$.

\section{Results}
\label{sec:results}

Results presented in this section concern matrices that obey transport-locality and lead to equilibrium-state.
Matrices that obey and violate strong-locality are constructed and discussed for both 
types of particles, distinguishable and indistinguishable ones.

Firstly
we introduce the microstate-symmetry used to classify matrices of the Symmetric Cell Models~(\ref{sec:ms}).
Then the  microstate-symmetric matrix for 
distinguishable particles~(\ref{sec:msm}) and 
the matrix for 
indistinguishable particles approximating 
the microstate-symmetric one~(\ref{sec:simplest})
are presented and discussed. 
Finally, examples of matrices for distinguishable and indistinguishable particles
having complementary properties concerning strong-locality are given~(\ref{sec:om}).

\subsection{Microstate-symmetry}
\label{sec:ms}

The strongest symmetry one can impose on the system is the microstate-symmetry.
It includes the previously introduced symmetries - the space-time symmetries and
the transition-matrix symmetry. 
To define the microstate-symmetry we introduce
\emph{allowed} and \emph{forbidden} transitions.
The allowed transitions are between microstates that fulfil constraints imposed on the system. For this work, the constrain of transport-locality is of relevance.
The forbidden transitions are not allowed transitions. Note that all transition matrix elements
corresponding to the forbidden transitions are zero, whereas the allowed transitions may
have non-zero and zero probabilities. The latter ones are due to selected dynamics of
the system evolution.

Let $M$ be an arbitrary transformation of microstates that preserves the set. That is the set of transformed microstates is the same as the original set of microstates.
In addition, let $M$ preserves the allowed and forbidden transitions, that is, 
if $X \to Y$ is  allowed~(forbidden) then $M(X)\to M(Y)$ is allowed~(forbidden) as well.
Then we call the transition matrix \textit{microstate-symmetric} if:

\begin{equation}
\label{eq:microstate_symmetry}
B(X\to Y|X) = B(~M(X)\to M(Y)|M(X)~)~, \qquad \forall M.
\end{equation}
It is immediate to check that a microstate-symmetric matrix, among the others,  respects translation, reflection and matrix symmetries.
From the microstate-symmetry follows that the probabilities of allowed transitions are equal:

\begin{equation}
\label{eq:microstate_symmetry_1}
B(X_i\to Y_j|X_i) = const
\end{equation}
for any $i, j$ leading to an allowed transition and vanishing otherwise.

Needless to say that in the case of all transitions being allowed, including teleportation ones, 
the microstate-symmetric matrix leads to the equilibrium state after the first time step independently of the assumed initial state. Thus the matrix has the fastest equilibration time, $t_{eq} = 1$.

\begin{figure}[htp]

\includegraphics[width=.40\textwidth]{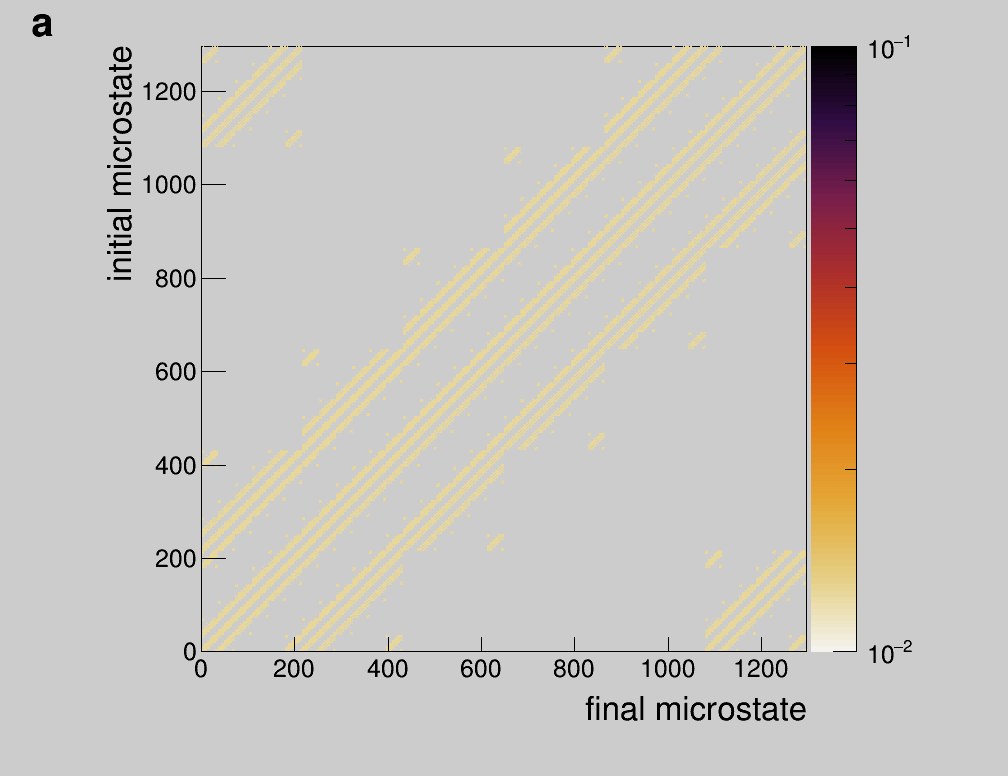}
\includegraphics[width=.40\textwidth]{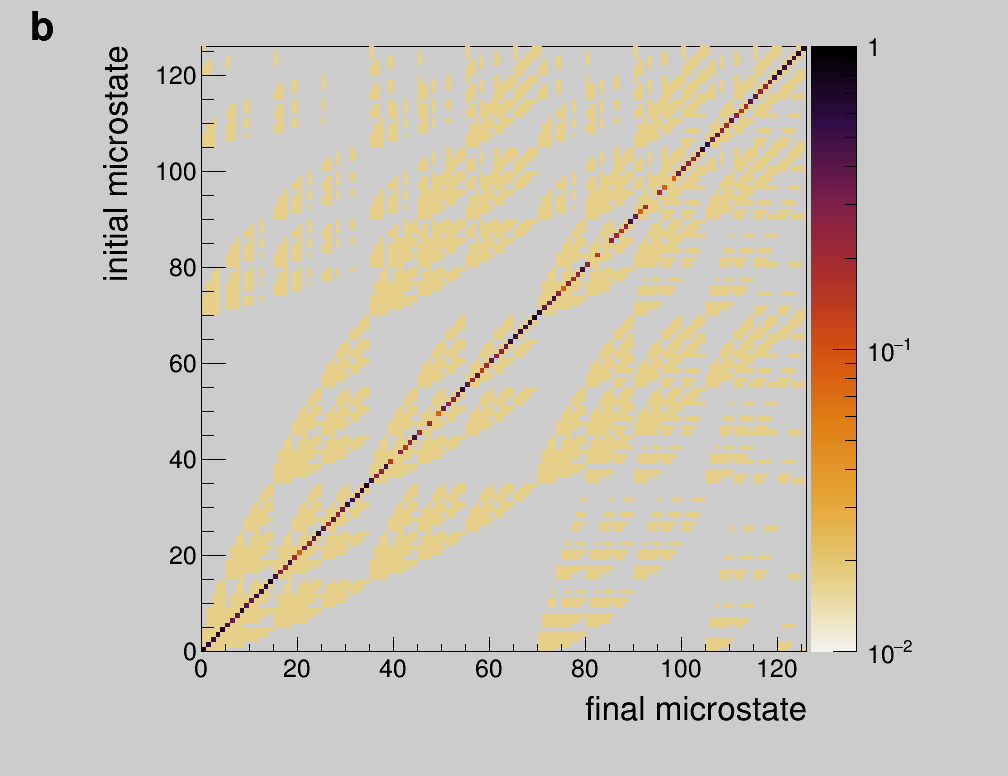}
\includegraphics[width=.40\textwidth]{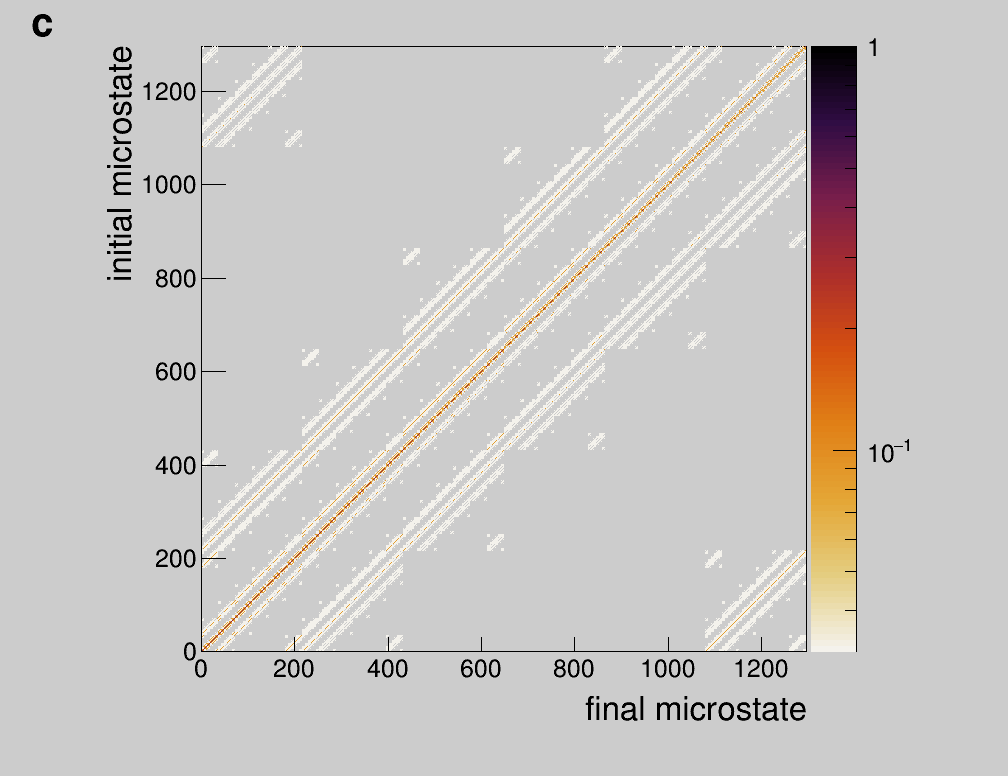}
\includegraphics[width=.40\textwidth]{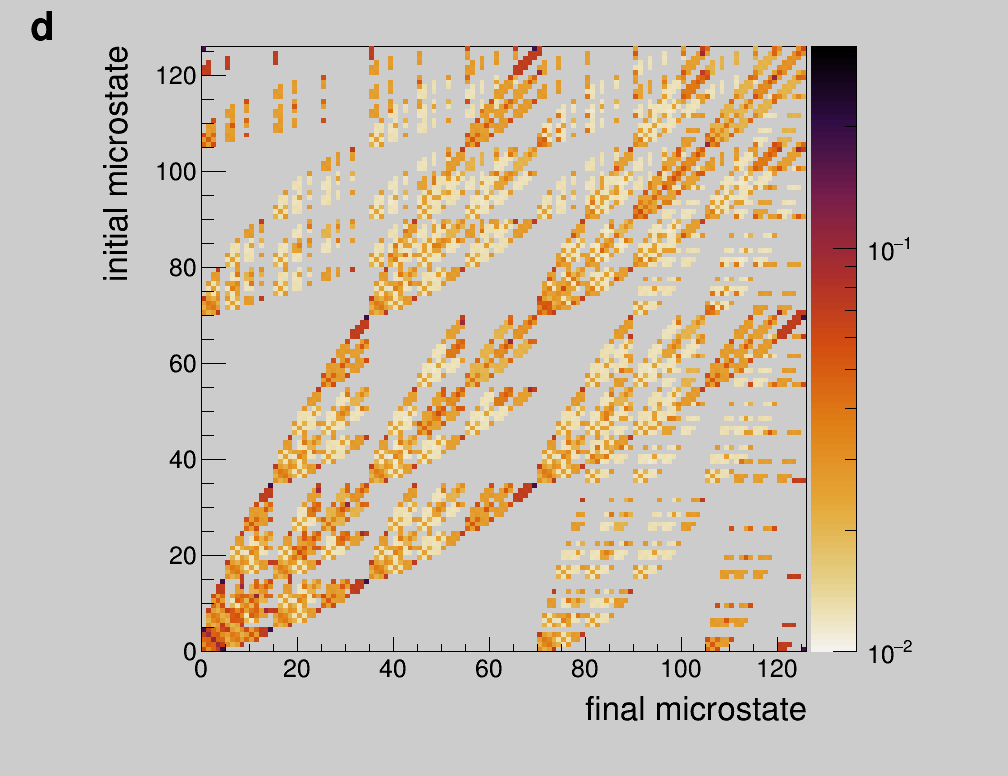}

\caption{
Transition matrices of the Symmetric Cell Models for four particles in six cells.
The matrices for distinguishable particles are shown in the left plots and the
 indistinguishable particles in the right plots.
All matrices obey transport-locality and lead to equilibrium-state.
The  microstate-symmetric matrix for distinguishable particles is shown 
in the plot (a).
It is of the type $\B\in$[EQ+~TL+~SL+].
The matrix approximating the microstate-symmetric one for indistinguishable particles is shown in the plot (b) and
it is of the type $B\in$[EQ+~TL+~SL-]. 
The bottom plots show the $\B\in$[EQ+~TL+~SL-] matrix for distinguishable particle (c) 
and the $B\in$[EQ+~TL+~SL+] for indistinguishable ones (d). 
Microstates are ordered according to their sequential number $S(X)$ defined as a position of the microstate  in a vector of ``human-friendly'' microstates labels 
calculated as  for $\sum_{i=1}^N  v_i \cdot 10^{i-1}$ and $\sum_{i=1}^{V} n_{V-i} \cdot 10^{i-1}$ for distinguishable  and indistinguishable particles, respectively.
The colour scale indicates transition probability, with the white colour denoting zero probability for allowed transitions. Transitions forbidden by the transport-locality are in grey. Note different colour scales adopted to underline qualitative differences between
the matrices.
}
\label{fig:transition_matrix}
\end{figure}

\subsection{Microstate-symmetric matrix for distinguishable particles}
\label{sec:msm}

It is easy to construct the transport-local matrix
obeying the microstate-symmetry for distinguishable particles.
This is done by setting to zero the probabilities of the transitions violating 
transport-locality and to a constant the remaining ones. Because of the normalisation to $1$ of the total probability, this constant must be equal to the inverse of the total number of transport-local transitions from a given initial microstate. It is immediate to check that, for distinguishable particles, this number does not depend on the initial microstate but only on the total number of particles $N$ and 
the maximum distance $\Delta$.
In addition to transport-locality, the transition matrix also preserves strong-locality. This is obvious,  noticing that the
matrix construction is equivalent to a model in which a distinguishable particle has an equal probability of appearing in any of the cells, which are at most $\Delta$ cells distant from the starting cell.
An example matrix of this type ($\B\in$[EQ$+$~TL$+$~SL$+$]) calculated for $V=6$,
$N=4$ and $\Delta=1$ is shown in Fig.~\ref{fig:transition_matrix}~(a).

\subsection{Approximating the microstate-symmetric matrix for indistinguishable particles}
\label{sec:simplest}

In the case of indistinguishable particles,
the transition matrix which  obeys microstate-symmetry does not exist.
It is impossible to set all probabilities of the transport-local transitions to a constant. In general, there are fewer microstates for indistinguishable particles than for distinguishable ones for the same multiplicities in cells.
And the number of transport-local transitions depends on the initial state. 
On the other hand, the probability normalisation condition $\sum_Y B(X\to Y|X) = 1$ is independent of the initial microstate. 
The latter two requirements are in contradiction.
The above is easy to see considering transitions from two examples of initial microstates for distinguishable and indistinguishable particles.
The first one have all particles distant from each other by more $2 \cdot \Delta$.
In the second example all particles are in the same cell.

The matrix for indistinguishable particles that most closely approximates
the microstate-symmetric matrix is constructed as follows.
All off-diagonal elements corresponding to transport-local transitions are equal and non-zero. The ones violating transport locality are set to zero.
Then diagonal elements are calculated from the probability normalisation condition. 
By construction, the matrix is symmetric and obeys all the space-time symmetries of the Symmetric Cell Models.
The matrix is also ergodic; see Appendix~\ref{app:a} for detail.
Thus, it belongs to the class of equilibrating Symmetric Cell Models.
By construction, the matrix obeys transport-locality. 
However, it violates strong-locality~(\ref{sec:sl}).
This is because the strong-locality condition ($\nu_{SL} = 0$) 
for indistinguishable particles requires  different 
off-diagonal elements for initial microstates with different numbers of transport-local
transition, whereas
the matrix has identical off-diagonal elements.
An example matrix of this type ($B\in$[EQ$+$~TL$+$~SL$-$]) calculated for $V=6$, 
$N=4$ and $\Delta=1$ is shown in Fig.~\ref{fig:transition_matrix}~(b). 
The  strong-locality violation parameter 
(\ref{eq:Rstrong})
for this matrix is $\nu_{SL} \approx 0.010$.

One can construct the microstate-symmetric matrix for indistinguishable particles by redefining the set of allowed transitions such that their number is the same for all microstates. This requires assuming that either not all transport-local transitions are allowed or teleportation transitions are possible.

\subsection{Complementary types of transition matrices}
\label{sec:om}

Here we construct transport-local and equilibrating matrices for distinguishable particles violating strong-locality ($\B\in$[EQ$+$~TL$+$~SL$-$])
and  for indistinguishable particles obeying
strong-locality ($\B\in$[EQ$+$~TL$+$~SL$+$]).
Concerning the strong-locality, they complement matrices discussed in 
Secs.~\ref{sec:msm} and~\ref{sec:simplest}.
The unusual properties of the matrices are obtained at the expense of their rather
complex structure, see Fig.~\ref{fig:transition_matrix}~(c) and~(d).

\vspace{0.5cm}
\noindent
\textbf{
$\B\in$[EQ$+$ TL$+$ SL$-$]}:
It is easy to construct transition matrices for distinguishable particles leading to an equilibrium-state
preserving transport-locality and violating strong-locality.
For example,
one assumes that only a single randomly selected particle can move to the right or left cell at each time step. 
It is straightforward to see that 
the matrix fulfils the requirements of the equilibrating Symmetric Cell Models.
By construction, the matrix obeys transport-locality. 
However, it violates a strong-locality - the movement of 
a particle from a given cell implies that all other particles do not move.
An example matrix of this type calculated for $V=6$, $N=4$ and $\Delta = 1$ 
and setting
probabilities to move to the right and left to $1/3$ is shown in Fig.~\ref{fig:transition_matrix}~(c).

\vspace{0.5cm}
\noindent
\textbf{
$B\in$[EQ$+$~TL$+$~SL$+$]}:
The construction of a matrix for indistinguishable particles that
leads to equilibrium preserving strong-locality is more complicated.
It is possible to enforce strong-locality starting from the already introduced 
procedure of independent cell-by-cell redistribution of cell multiplicities
(see Sec.~\ref{sec:equilibrium} and Appendix~\ref{app:d} on the $B\in$[EQ$-$~TL$+$~SL$+$]). 
One takes particle multiplicity $n$ in a cell, and for each of its 
redistribution in the neighbouring cells (no more than $\Delta$ cells away) 
assigns a probability. 
The final microstate is obtained by drawing one of the redistributions for each cell and
summing over contributions from all cells.
By assumption of the independent cell-by-cell redistribution within
$\pm\Delta$ cells, the model obeys transport- and strong-locality.
It is straightforward to select probabilities that obey space-translation 
and reflection symmetries - the resulting matrix is ergodic. 
To assure the symmetry of the transition matrix is less trivial. 
Assuming the equiprobability of each of the redistribution of $n$ particles in the neighbouring cells, one constructs the matrix fulfilling the required constraints besides the matrix-symmetry (see Appendix~\ref{app:c}). It is, however, possible to enforce the symmetry of the transition matrix, starting with the equiprobable redistribution for $n=1$ and enforcing symmetry for $n>1$ in a 
recursive procedure. The matrix-symmetry requires that some redistributions for $n$ particles have selected probabilities fixed by cases with less than $n$ particles. The construction is rather technical and presented in Appendix~\ref{app:d} in detail. 
Its validity was checked up to $N=1000$.
An example of a matrix of the type $B\in$[EQ$+$~TL$+$~SL$+$], calculated for $V=6$, $N=4$ and $\Delta=1$ is shown in Fig.~\ref{fig:transition_matrix}~(d).

\subsection{Summary}

We consider here previously unasked question - how does   
evolution of an assembly of particles to the equilibrium-state   
relate to strong-locality?
More specifically, whether, concerning this question, indistinguishable particles differ from distinguishable ones.
The question is motivated by quantum mechanics, which encompasses both
properties, equilibration and violation of strong-locality, 
for indistinguishable particles.

The framework introduced here to address the question 
is based on discrete-time Markov chains over a finite set of microstates of
indistinguishable and distinguishable particles.
The space-time symmetries are imposed. 
For the first time, we introduce conditions needed to
obey the
transport- and strong-locality for indistinguishable particles 
within Markov chains.
The constraints and dynamics are encapsulated in the Markov-chain transition matrix defining the time evolution of the assembly of particles.

Models which obey particle transport-locality 
and lead to the equilibrium-state are considered.
We show that it is possible to construct models obeying and violating
strong-locality both for indistinguishable and for distinguishable particles.
To uncover differences between the two types of particles, 
we introduce the microstate-symmetry
- equiprobability of allowed transitions. 
It is the strongest symmetry, and it includes the  symmetries considered in this work - the space-time and matrix symmetries.

For distinguishable particles,
the microstate-symmetric model exists.
It leads to equilibrium-state obeying strong-locality.
For indistinguishable particles, the microstate-symmetric model does not exist. 
The matrix for indistinguishable particles which most closely approximates
the microstate-symmetric one violates strong-locality when evolving to
the equilibrium-state. These findings suggest an answer to the primary question -
only for distinguishable particles, the strongly-local evolution to equilibrium
is possible without breaking the microstate-symmetry.

Depending on postulates of a model,
the answer may lead to different consequences.
For example, assuming the that dynamics of indistinguishable particles fulfils the microstate-symmetry, one concludes that either not all transport-local transitions are allowed or teleportation transitions are possible. One may speculate that this and other conjectures inspired by the results presented here can be tested by studying high-energy
nuclear collisions. They indicate that the apparent equilibrium hadronic-final state of many indistinguishable particles
is produced in collisions within a short time~\cite{Florkowski:2010zz}. 

We hope that the work provides new tools and perspectives on strong-locality and its violation.

\section{Code and numerical results availability}
Code reproducing the key results of this paper is available under the link \href{https://gitlab.cern.ch/ipidhurs/cell-model-fitter}{https://gitlab.cern.ch/ipidhurs/cell-model-fitter}.
Under the same link, one will also find numerical values for the matrices presented in the plots.

\clearpage

\begin{acknowledgments} 
We are thankful to F.~Giacosa, B.I. Lev, St. Mrowczynski, and A.A. Semenov for their comments, and 
M.~Gazdzicka and M.~Prior for help in the preparation of the manuscript.
This work is partially supported by
the Polish National Science Centre grant 2018/30/A/ST2/00226, 2016/21/D/ST2/01983, 2020/39/D/ST2/02054, the German Research Foundation grant GA1480\slash 8-1, Collaborative Research Center CRC-TR 211 project number 315477589 - TRR 211, and
by the National Academy of Sciences of Ukraine, Grant No. 0122U200259.
\end{acknowledgments}

\clearpage


\appendix
\section{Ergodicity of the discussed  matrices}
\label{app:a}

Here we show that all matrices discussed in this work are ergodic.

We recall (Sec.~II.B)
the requirements for ergodicity in a finite Markov chain~\cite{ergodic}:
\begin{enumerate}[(a)]
 \item
 \emph{Irreducibility}. Each microstate is reachable from any other microstate by a sequence of transitions with non-vanishing probability.
 \item
 \emph{Aperiodicity}. 
 The maximum common divisor of the number of transitions of all sequences linking one microstate to itself is one.
\end{enumerate}

The discussed matrices are irreducible because, at the very least, transitions that change only multiplicities of any two adjacent cells (one or both being non-empty)  by one unit have a non-zero probability.
Consequently, by repeating these transitions, one can reach any microstate starting from any initial microstate.

The matrices are aperiodic because, at the very least, one of the transitions which do not change a microstate
($X \to X$) have non-zero probabilities in all matrices.
Consequently,
any number of time steps for a microstate to recur can be increased by one step
by adding the transition  $X \to X$.

\section{Example of the [EQ-~TL+~SL+] matrix for indistinguishable particles}
\label{app:b}

It is easy to construct a transport- and strong-local Symmetric Cell Model~(Sec.~II.B) having  a non-equilibrium steady-state. An interesting example is given below.

One assumes that during a transition, the number of particles in the cell $i$ at $t$ 
is redistributed between the cells from $i-\Delta$ to $i+\Delta$ at $t+1$, such as that all possible configurations 
have  equal probability. That is, all the (partial) redistributions have a probability equal to the inverse of the total number of  redistributions of $n$ particles within the $2\Delta+1$ cells:

$$
\frac{(2\Delta)! n!}{(n+2\Delta)!}~.
$$
The  redistribution of particles is performed independently for all initial microstate cells. 
Then the final microstate is constructed by summing the partial contributions to each cell
of the final microstate from all cells of the initial microstate. 
The transition probability is the product of the probabilities of each partial redistribution in the single-cell times the number of different (total) redistributions reaching the same final microstate.

By construction - the particle number is redistributed locally and independently cell-by-cell -
the matrix obeys transport- and strong-localities,
$\mathcal{V}_{SL} = 0$.

 The translation and reflection symmetries are  simple to check since the partial redistributions are equiprobable (no asymmetry between particles "going to the right or the left"). The redistribution probabilities are independent of the cell number.
 
 The ergodicity can be proven with the general arguments presented  above  
 (Appendix~\ref{app:a}), which are valid for all matrices considered in the paper, 
 including this one. 
 
 With the transition matrix, it is possible to compute the steady-state, which appears to be  non-equilibrium. In fact, in the next section, we show proof that the transition matrix is time-reversible but not symmetric.

\begin{figure}[ht!]
    \centering
    \includegraphics[width=.49\textwidth]{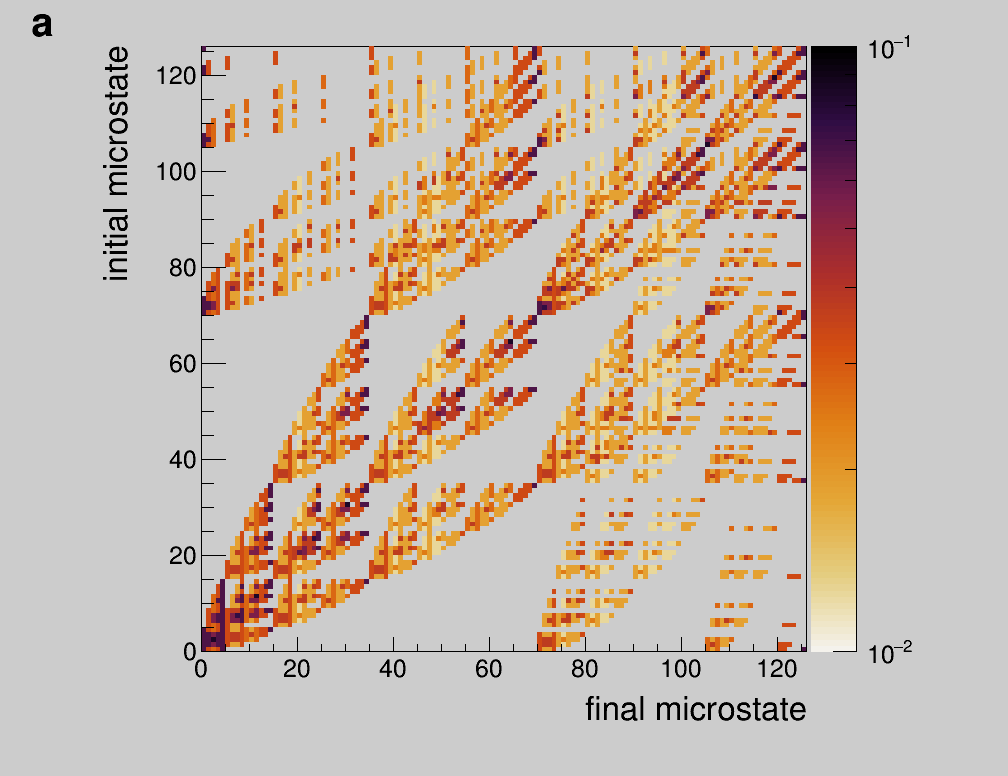}
    \includegraphics[width=.49\textwidth]{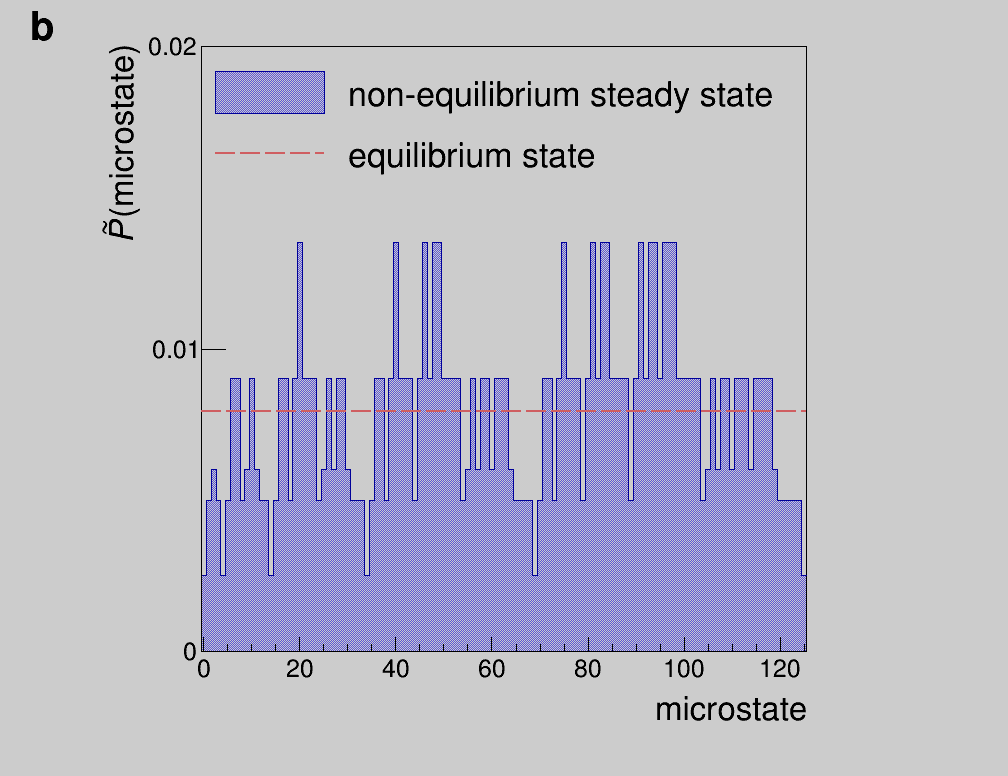}
    \caption{
        (a):
Example of the $B\in$[EQ-,~TL+,~TL-] transition matrix for four indistinguishable particles in six cells. The matrix  
leads to non-equilibrium steady-state obeying transport- and 
strong-locality.  
Transitions forbidden by the transport-locality are in grey.
(b):
The microstate probabilities in the steady-state of the example matrix  
(in blue). The red dashed line corresponds to the probabilities at equilibrium, which are by definition equal to $1/W_{ind}$, where $W_{ind}$ is the total number of microstates. \\ 
Microstates are ordered according to their
sequential number $S(X)$ defined as a position of the microstate  in a vector of ``human-friendly'' microstates' labels 
calculated as  $\sum_{i=1}^{V} n_{V-i} \cdot 10^{i-1}$.
        }
    \label{fig_m:transition_matrix}
\end{figure}

Thus, the assumption of statistical and local (cell-by-cell) redistribution of cell multiplicities, unlike naively expected,
leads to a non-equilibrium (non-statistical) steady-state.
The model transition matrix is  $B\in$[EQ$-$~TL$+$~SL$+$]. 
The matrix  calculated for $V=6$, $N=4$ and $\Delta = 1$ is shown in Fig.~\ref{fig_m:transition_matrix}~(a) and  the corresponding microstate
probability distribution of the steady-state in Fig.~\ref{fig_m:transition_matrix}~(b).

\section{Asymmetry of the example [EQ-~TL+~SL+] matrix for indistinguishable particles}
\label{app:c}

Here we discuss  the steady-state of
the transition matrix $B\in$[EQ$-$~TL$+$~SL$+$] introduced in Appendix~\ref{app:b}. 
Doing so we show that the transition matrix is time reversible but not symmetric.
Its example for $V = 6$, $N = 4$ and $\Delta = 1$ is presented in Fig.~\ref{fig_m:transition_matrix} (a). 

By construction

\eq{\label{b_const}
 B(X\to Y|X) = \frac{s(X\to Y)}{r(X)}~,
}
where $r(X)$ is the total number of partial redistributions from the microstate $X=(n_1^X,\cdots, n_V^X )$

\eq{
r(X)=\prod_{i=1}^V \frac{(n_i+2\Delta)!}{(2\Delta)!n_i!}~, 
}
and $s(X-Y)$ is the total number of different redistributions going from $X$ to $Y$. 

Note that $s(X\to Y)$ is symmetric $s(X\to Y)=s(Y\to X)$. For each partial redistribution of the particles from $X$ to $Y$, there is a canonical partial redistribution from $Y$ to $X$ - it is the same process in the reverse direction. 
Of course, in general $r(X)\neq r(Y)$, as we will discuss below.

From the symmetry of $s$ \eqref{b_const} one gets

\eq{
B(X\to Y|X)= \frac{s(X\to Y)}{r(X)} = r(Y)\frac{s(Y\to X)}{r(Y)r(X)} = \frac{r(Y)}{r(X)} B(Y\to X|Y)~.
}
This is equivalent to

\eq{
r(X) B(X\to Y|X) = r(Y) B(Y\to X|Y)~.
}
At this point, it is simple to write the steady-state distribution 
$\pi_S(X)$, which is just the $r(X)$ normalized to one

\eq{
\pi_S(X) = \frac{r(X)}{\sum_Z r(Z)}~, 
\qquad \pi_S(X) B(X\to Y|X) = \pi_S(Y) B(Y\to X|Y)~.
} 
 Clearly $\pi_S(X)$ is time reversible if it is the steady-state. It can be easily proved that it fulfills the definition of 
 the steady-state by summing over $X$, reminding that $\sum_X B(Y\to X|Y)=1$.
 
 The transition probability matrix $B(X\to Y|X)$ is thus ergodic and time-reversible, but not symmetric. Therefore,  the steady state $\pi_S(X)$ cannot be the equiprobable one. There must be $X$ and $Y$ with $r(X)\neq r(Y)$. To verify the latter, 
 it is enough to note that for $N>1$ there is always at least one transition with $B(X\to Y|X)\neq B(Y\to X|Y)$.
Let us consider  the initial microstate $X$ with $n_i^{X} = N$, and the final one $Y$ with $n_i^{Y} = N-1$ and $n_{i+1}^{Y} =1$. From Eq.~\eqref{b_const} follows that the probability $B(X\to Y|X)$
is equal $(2\Delta)!N!/(N+2\Delta)!$.
On the other hand, the probability of the inverse transition, $B(Y \to X|Y)$, is $1/(2\Delta+1)$ (the partial probability of the particle in cell $i+1$ to move to the cell $i$) times $(2\Delta)!(N -1)!/(N+ 2\Delta -1)!$ (the probability that $N-1$ particles in cell $i$ remain in cell $i$). 
Consequently

\eq{\begin{split}
\frac{1}{2\Delta+1}\frac{(2\Delta)!(N-1)!}{(N+2\Delta-1)!}= \frac{1}{2\Delta +1}\frac{N+2\Delta}{N} \frac{(2\Delta)!N!}{(N+2\Delta)!} &= B(Y \to X|Y) \\
= B(Y \to X|Y) <  B(X \to Y|X) = \frac{(2\Delta)!N!}{(N+2\Delta)!}~.
\end{split}
}
In particular, in the large $N$ limit, the transition $X \to Y$ has a probability $2\Delta +1$ larger than the transition $Y \to X$.

\section{Construction of the [EQ+~TL+~SL+]  matrix for indistinguishable particles}
\label{app:d}

Here we explicitly construct
the transition matrix $B\in$[EQ$+$~TL$+$~SL$+$] discussed in Sec.~III.D.
It obeys 
transport- and strong-localities (Sec.~II.C), and it is symmetric and ergodic.

The construction is based on the procedure presented in Appendix~\ref{app:b} for the $B\in$[EQ$-$~TL$+$~SL$+$] matrix. However, the probability of partial redistributions is not assumed to be constant. The set of partial redistributions for a cell with $n$ particles is obtained using the recursive procedure starting from the system with $n = 1$ particles and fixing the probabilities for the higher $n$ enforcing symmetry order by order. The ergodicity and the  symmetries are obeyed, as long as the probabilities to "move left" are equal to the ones "moving right".

For simplicity, we will consider the case of $\Delta =1$. It is straightforward to generalize the procedure for an arbitrary $\Delta$.

Possible (partial) redistributions of a single particle in cell $i$ at $t$ to move the neighbouring cells $\{n_{i-1},n_i,n_{i+1}\}$ at $t+1$ are $\{1,0,0\}$, $\{0,1,0\}$ and $\{0,0,1\}$. 
Assuming probabilities to 
"move right" ${\sf P}^{(1)}_R$, "move left"  ${\sf P}^{(1)}_L$  and stay ${\sf P}^{(1)}_S$ in the cell $i$ 
are equal, ${\sf P}^{(1)}_R={\sf P}^{(1)}_L={\sf P}^{(1)}_S = 1/3$,
one gets the simplest case. It is worth mentioning however, that one can generalize the procedure as long as the probability to "move righ" and "move left" are equal.

For $n = 2$, there are two cases:
\begin{enumerate}[(i)]
    \item
    Two particles in the same cell "split" into different cells. We fix the probability of the partial redistribution to be equal to the probability of the partial redistributions for the inverse process ${\sf P}^{(2)}_{SPLIT}=({\sf P}^{(1)}_{SPLIT})^2=1/9$. There are three possible outcomes, depending on which cell remains unoccupied in the partial redistribution.
    \item
    Two particles in the same cell "stay together" in the processes. There are still three outcomes (the cell in which both particles land), we fix them assuming that they are equiprobable, and using the fact that the sum of all probabilities must be $1$
    \eq{\label{eq:stay_2}
    {\sf P}^{(2)}_{S} = \frac{1} {3} \left[ 1 - {\sf P}^{(2)}_{SPLIT} \right] = 2/9~.
    }
\end{enumerate}
Thus, by construction, all the partial redistributions of cell multiplicities
have probabilities equal to the corresponding reverse processes. The above procedure must be repeated recursively up to $N$. It grants that the  transition matrix is symmetric. The number of independent redistributions $s(X\to Y)$ is independent of the redistribution probabilities and symmetric $s(X\to Y) = s(Y\to X)$. 
The probability of each independent process connecting $X$ to $Y$ has its inverse that by construction has the same probability. 
Thus the probability of each partial redistribution has the same numerical value as its inverse. 

Then the "stay-together" probabilities for $n>2$ up to $N$ depends on the total number of available partial transitions and the probabilities obtained for smaller $n$. 
They can be shown to   be equal to
\begin{equation}
\label{eq:stay_N}
 {\sf P}^{(n)}_S= \frac{1}{3}\left[ 1-\sum_{0<k\le l\le n-k-l} {\sf C}_{n-k-l,k,l}{\sf P}_S^{(n-k-l)}{\sf P}_S^{(k)}{\sf P}_S^{(l)} - \sum_{0<m\le n-m}{\sf C}_{n_i-m, m}{\sf P}_S^{(n-m)}{\sf P}_S^{(m)}\right]~,
\end{equation}
where the number of different  combinations ${\sf C}_{n-k-l,k,l}$ of particles in three cells with multiplicities $n-k-l$, $k$ and $l$  is either $1$ (if $k=l=n-k-l$), $3$ (if either $k=l< n-k-l$ or $k<l=n_i-k-l$), or $6$ (if $k<l<n-k-l$). In the same way the combinations ${\sf C}_{n-m,m}$ are either 6 (if $n-m>m$) or $3$ (if $n-m = m$).
We have no proof that ${\sf P}^{(N)}_S$ remains positive, and thus
the transition matrix exists for arbitrary large $N$. However, we checked numerically that Eq.~\ref{eq:stay_N} gives ${\sf P}^{(n)}_S > 0$ for $n$ up to 1000. 
It is possible to verify the construction for any finite $N$ providing a sufficient
computing time is available.

\newpage
\bibliographystyle{ieeetr}
\bibliography{references}

\end{document}